\documentclass[aps,prd,superscriptaddress,twocolumn,showpacs]{revtex4-1}
\usepackage{graphicx}
\usepackage{dcolumn}
\usepackage{color}  
\usepackage{bm}

\newcommand{\ls}{\lower0.5ex\hbox{$\; \buildrel < \over \sim \;$}}
\begin{document}

\title{Properties of Bare Strange Stars Associated with Surface Electric Fields}

\date{\today}

\author{Rodrigo Pican\c{c}o Negreiros} 
\affiliation{Frankfurt Institute for Advanced Studies, Goethe
  University, Ruth Moufang Str.\ 1, 60438 Frankfurt am Main, Germany}

\author{Igor N.\ Mishustin} 
\affiliation{Frankfurt Institute for Advanced Studies, Goethe
   University, Ruth Moufang Str.\ 1, 60438 Frankfurt am Main, Germany}
\affiliation{The Kurchatov Institute, Russian Research Center,
    123182 Moscow, Russia}

\author{Stefan Schramm} 
\affiliation{Frankfurt Institute for Advanced Studies, Goethe
  University, Ruth Moufang Str.\ 1, 60438 Frankfurt am Main, Germany}
\affiliation{CSC, ITP, Goethe University, Frankfurt Institute
    for Advanced Studies, Goethe University, Ruth Moufang Str.\ 1,
    60438 Frankfurt am Main, Germany Frankfurt am Main, Germany}

\author{Fridolin Weber} 
\affiliation{Department of Physics, San Diego State University, 5500
  Campanile Drive, San Diego, CA 92182, USA}

\begin{abstract}
  In this paper we investigate the electrodynamic surface properties
  of bare strange quark stars. The surfaces of such objects are
  characterized by the formation of ultra-high electric surface fields
  which might be as high as $\sim 10^{19}$~V/cm.  These fields
  result from the formation of electric dipole layers at the stellar
  surfaces.  We calculate the increase in gravitational mass
    associated with the energy stored in the electric dipole field,
    which turns out to be only significant if the star possesses a
    sufficiently strong {\em net} electric charge distribution. In
    part two of the paper, we explore the intriguing possibility of
    what happens when the electron layer (sphere) rotates with respect
    to the stellar strange matter body. We find that in this event
    magnetic fields can be generated which, for moderate effective
    rotational frequencies between the electron layer and the stellar
    body, agree with the magnetic fields inferred for several Central
    Compact Objects (CCOs). These objects could thus be comfortably
    interpreted as strange stars whose electron atmospheres rotate at
    frequencies that are moderately different ($\sim 10$~Hz) from the
    rotational frequencies of the strange star itself.
\end{abstract}

\maketitle

\section{Introduction}

The properties of hypothetical strange quark stars (strange stars for
short) have been extensively investigated in the literature
\cite{Alcock1986,Alcock1988,Weber2005,Madsen1999,Glendenning1992a,
Glendenning1995}. Strange stars are
compact astrophysical objects composed by absolutely stable strange
quark matter \cite{Witten1984,Bodmer1971}, which has a zero-pressure point at
finite baryon density $n_* \geq n_0$, where $n_0 = 0.15$fm$^{-3}$ is the
equilibrium nuclear density. This matter consists of a roughly
equal numbers of up, down and strange quarks and a relatively small
number of electrons. Toward the quark star surface, however, the
number of strange quarks drops and a higher number of electrons is
required to maintain electric charge neutrality
\cite{Alcock1986,Alcock1988,Weber2005,Madsen1999}. The electrons, not
being bound by the strong interaction, are displaced to the outside of
the star and an electric dipole layer is formed at the surface
of a quark star \cite{Alcock1986,Usov2004,Kettner1995}. In the region between
the
core surface and the electron layer a static electric field of the
order of $10^{17-19}$ V/cm is formed \cite{Alcock1986,Usov2004,Kettner1995}. The
actual value of the electric field will depend on electrostatic
effects, including Debye screening, the surface tension of the
interface between the vacuum and quark matter
\cite{Alford2006,Jaikumar2006}, on whether or not the matter is in a
superconducting state
\cite{Rajagopal2001,Alford2001,Linares2006,Rajagopal2001a,Alford2002},
and on the degree of sharpness of the surface
\cite{Mishustin2010}. This electric field may support a thin nuclear
crust (thickness $\sim 100$~m,
\cite{Alcock1986,Glendenning1992,Glendenning1995,Weber2005})
consisting of a lattice of ordinary atomic nuclei below neutron drip
density. Alternatively, strange stars may also lack a nuclear crust in
which case they are referred to as bare strange stars. In this paper
we will consider the latter possibility.

In the first part of this paper we will focus on the electrostatic
properties of strange stars, treated within the framework of general
relativity. We extend the results presented in
\cite{Negreiros2009,Ray2003}, where the general relativistic structure
of compact stars with a net electric surface charge was analyzed. As
shown there, the energy density associated with the electric surface
field acts as a source of curvature and thus contributes to the
gravitational masses of strange stars. In the present paper we extend
the results of \cite{Negreiros2009} to include the electron layer. We
will show that even  under the most extreme physical conditions
  possible, the energy density associated with the electric dipole
layer at the surface of a strange star does not lead to a substantial
change of the star's gravitational mass. Only when the star possesses
a net electric charge, as considered in \cite{Negreiros2009}, one
obtains a non-negligible increase in the gravitational mass as a
result of the electric field.

In the second part of the paper we consider the consequences of stellar
rotation for the electric fields. We allow the strange star and the
electron layer surrounding it to rotate at different
frequencies. Under these circumstances, electric currents will exist
at the surface of the star. The associated magnetic field is
calculated. This field turns out to be uniform inside the star but
dipolar outside of the star. It is shown that for a certain range of
frequencies and electrostatic properties, the computed magnetic field
is in agreement with those  inferred for three Central Compact
Objects (CCOs) (\cite{Becker1932,Halpern2010} and references
therein). The emission from CCOs is characterized by a steady photon
flux in the X-ray range and the absence of emission radio and optical
counterparts. As shown in \cite{Halpern2010}, several of these objects
are slowly rotating (see Table \ref{table:CCO}) and possess relatively
low magnetic fields. The findings of this paper could indicate that
these CCOs are not neutron stars but rather  strange stars whose
  surrounding electron layers rotate slowly relative to the strange
  matter cores.  There is evidence that CCOs have unusually small
projected emitting  areas, in the range of $0.3 -5 $~km
\cite{Becker1932}, which, if confirmed, would support the
interpretation of CCOs as strange stars.

This paper is organized as follows. In Section \ref{section:II} we
review the structure equations of electrically charged compact
stars. The change of the gravitational mass by ultra-strong electric
dipole fields on quark stars is computed in Section
\ref{section:III}. This is followed in Section \ref{section:IV} by an
investigation of the magnetic surface properties of strange stars.  A
summary and conclusions are provided in Section
\ref{section:conclusions}.

\section{Structure of Electrically Charged Compact Stars}
\label{section:II}

Electrically charged compact stars are described by an energy-momentum
tensor of the following type
\cite{Bekenstein1971,Negreiros2009,Ray2003},
\begin{eqnarray}
  T_{\nu}^{\mu} &=& (P +\epsilon \, c^2)u_{\nu} u^{\mu} + 
  P \, \delta_{\nu}^{ \mu}
  \nonumber \\
  &&+\frac{1}{4\pi} \left( F^{\mu l} F_{\nu l} -\frac{1}{4}
    \delta_{\nu}^{\mu} F_{kl} F^{kl} \right) \, ,
\label{eq:tnumu}
\end{eqnarray}
where $F^{\nu \mu}$ is the electromagnetic field tensor which 
enters the Maxwell equations as
\begin{equation} [(-g)^{1/2} F^{\nu \mu}]_{, \mu} = 4\pi j^{\nu}
  (-g)^{1/2} \, .
  \label{ecem}
\end{equation}
The quantity $j^{\nu}$ is the electromagnetic four-current and $g
\equiv \mbox{det}(g_{\nu \mu})$ with the metric tensor given by
\begin{equation}
g_{\nu\mu} =\left( \begin{array}{cccc}
- e^{2\Phi(r)}     & 0 & 0 & 0 \\
0 & e^{2\Lambda(r)} & 0 & 0 \\
0 & 0 & r^2  & 0 \\
0 & 0 & 0 & r^2 \sin^2\theta
\end{array} \right) , \label{gnumu}
\end{equation}
For static stellar configurations, as considered in this paper, the
only non-vanishing component of the four-current is $j^0$.  Because of
symmetry reasons, the four-current is only a function of radial
distance, $r$, and all components of the electromagnetic field tensor
vanish, with the exception of $F^{01}$ and $F^{10}$, which describe
the radial component of the electric field. Upon writing the
energy-momentum tensor (\ref{eq:tnumu}) as
\begin{small}
\begin{equation}
T_{\nu}^{\mu} =\left( \begin{array}{cccc}
-\left( \epsilon + \frac{Q^2}{8\pi r^4} \right)  & 0 & 0 & 0 \\
0 & P - \frac{Q^2}{8\pi r^4} & 0 & 0 \\
0 & 0 & P + \frac{Q^2}{8\pi r^4}  & 0 \\
0 & 0 & 0 & P  +\frac{Q^2}{8\pi r^4}
\end{array} \right) \, , \label{TEMch}
\end{equation}
\end{small}
and substituting this expression into Einstein's field equation,
$G^\mu_\nu = (8 \pi/c^2) T^\mu_\nu$, one finds for the pressure gradient
inside electrically charged stars 
\begin{eqnarray}
  \frac{dP}{dr}  & = & - \frac{G\left( m +\frac{4\pi r^3}{c^2}
      \left( P - \frac{Q^{2} }{4\pi r^{4} c^{2}} \right) \right)}{c^{2} r^{2}
    \left( 1 - \frac{2Gm}{c^{2} r} + \frac{G Q^{2}}{r^{2} c^{4}} \right)}
  \ (P +\epsilon)\nonumber \\ & & +\frac{Q}{4 \pi r^4}\frac{dQ}{dr} \, ,
\label{TOVca} 
\end{eqnarray}
and for their gravitational masses
\begin{equation}
  \frac{dm}{dr} = \frac{4\pi r^2}{c^{2}} \epsilon
  +\frac{Q}{c^2 r}\frac{dQ}{dr} \, , \label{dmel}
\end{equation}
The quantity $Q$ denotes the electric charge located inside a region
of radius $r$ and is given by Gauss' law,
\begin{equation}
  Q(r) = 4\pi \int_{0}^{r}  r'^2 \; \rho_{\rm ch}(r') \; 
  e^{\Lambda(r')}  dr' \, .
  \label{Q}
\end{equation}
The quantities $\rho_{\rm ch}$ and $\Lambda$ in Eq.\ (\ref{Q}) denote
the local electric charge distribution and the star's radial metric
function, respectively \cite{Negreiros2009}.  Equations
(\ref{TOVca}) and (\ref{dmel}) are generalizations of the standard
Tolman-Oppenheimer-Volkoff (TOV) equation which describe the global
properties of electrically uncharged compact stars in the framework of
general relativity theory.  The standard Tolman-Oppenheimer-Volkoff
equation is obtained from Eqs.\ (\ref{TOVca}) and (\ref{dmel}) in the
limit $Q \rightarrow 0$. The occurrence of the $Q$ dependent terms in
Eqs.\ (\ref{TOVca}) and (\ref{dmel}) account for the Coulomb
interaction among the charged particles that are part of the star as
well as for the extra curvature produced by the energy density of the
electric field associated with the charges
\cite{Bekenstein1971,Negreiros2009,Ray2003}. In contrast to the
Coulomb interaction, which can be either attractive or repulsive,
depending on the electric charge, the additional contribution to
curvature increases the pull of gravity on the stellar matter.

As discussed in \cite{Negreiros2009}, electric fields are only
relevant for the structure of strange stars if $Q^2/8\pi r^4 \sim P$
(see also Eq.\ (\ref{TEMch})). This implies electric fields that are
on the order of $E \sim 10^{19 - 20}$~V/cm. Electric fields of this
magnitude, located at the surface of strange stars, increases
their gravitational masses and radii by up to 15\% and 5\%
respectively \cite{Negreiros2009}.


\section{Stellar Mass}\label{section:III}

We now turn our attention to the stellar mass equation
(\ref{dmel}). Integrating this equation leads to (making $c=1$)
\begin{equation}
  m(r) = 4 \pi \int_0^r\,  r'^2 \, \epsilon(r') dr' + m_{\cal E} \, ,
\label{m_int}
\end{equation}
where
\begin{equation}
 m_{\cal{E}} = \int_0^r \, \frac{Q(r')^2}{2 r'^2} \, dr' +
\frac{Q(r)^2}{2r} \, . \label{m_el_int}
\end{equation}
The first term on the right-hand-side of Eq.\ (\ref{m_int}) is the
standard expression for the gravitational mass of electrically
uncharged stars. The second term on the right-hand-side, $m_{\cal E}$,
accounts for the mass increase due to electric charges. This increase
consists of two contributions, the first one, $ \int_0^r (Q(r')^2/2
 r'^2) dr'$, represents the local mass-energy of the electric field
inside the star, and the second, $Q(r)^2/2r$, represents the total
mass-energy required to assemble the charges on the stellar
configuration. An object with global charge neutrality must have $Q(R)
= 0$, where $R$ is the stellar radius.  This implies that any increase in
the stellar mass originates from the energy density of local electric
fields.

\begin{figure}
\centering
\vspace{1.0cm}
\includegraphics[width=8.5cm]{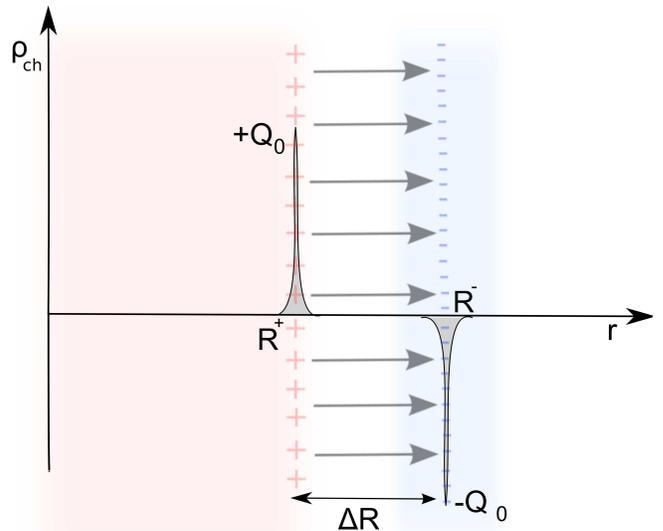}
\caption{\label{charge_scheme}(color online) Schematic representation
  of the electric charge distribution on the surface of a bare strange
  star. The core surface ($R^+$) becomes positively charged as the
  electrons ($R^-$) extend beyond the star's surface, giving rise to
  an electric dipole layer of width $\sim 10^3$~fm
  \cite{Alcock1986,Alcock1988,Weber2005,Usov2004,Mishustin2010}.}
\end{figure}

As discussed in references
\cite{Alcock1986,Alcock1988,Weber2005,Mishustin2010}, the surface of a
bare strange star consists of a positively charged layer surrounded by
an electron layer immediately outside of the star. This feature is
caused by the diminishing quark chemical potential toward the stellar
surface, which renders the (negatively charged) strange quarks less
abundant. This, in turn, requires a higher number of electrons (to
achieve global charge neutrality). Since the electrons do not feel the
the strong interaction that binds a strange star, they extend beyond
the stellar surface and form an electric dipole layer, with a
positively charged stellar surface.  Thomas-Fermi calculations
\cite{Alcock1986,Alcock1988,Usov2004,Mishustin2010} indicate that such
a configuration will produce electric fields of up to $10^{18}$ V/cm,
and a dipole region that extends $\sim 10^3$~fm beyond the star's
surface. We now investigate, in the framework of general relativity,
under which conditions such a configuration will have a significant
impact on the stellar mass. In order to do that, first we need to make
a suitable ansatz for the electric charge distribution ($\rho_{\rm
  ch}$). As mentioned just above, the dipole layer is of the order of
$\sim 10^3$ fm. Therefore, on a macroscopic scale we can safely assume
that both the positive and negative electric layers are described by
delta functions.  Adopting spherical coordinates, we have
\begin{equation}
  \rho_{\rm ch} = +K ~ \frac{\delta(r-R^+)}{4\pi r^2} - K ~ \frac{\delta(r -
    R^-)}{4\pi r^2} \, . \label{rhoch}
\end{equation}
Such charge distribution is shown schematically in Fig.\
\ref{charge_scheme}.
Integrating Eq.\ (\ref{Q}), with $\rho_{\rm ch}$ given by (\ref{rhoch}),
we obtain
\begin{equation}
 Q(r) = \Bigg \{ \begin{tabular}{cl}
         $+Q_0$  &for  $R^+ ~ \leq r ~ \leq R^-$\, ,  \\
                 & \\
         $0$     &elsewhere \, , \\
        \end{tabular} 
\end{equation}
where $Q_0 = K\times \exp(\Lambda(R))$.  To obtain this result, we
made use of the fact that the metric is essentially constant over the
relevant region, $\sim 10^3$~fm at the star's surface, $R$.  The
electrostatic contribution to the mass of the star can then be written as
\begin{equation}
  M_{\cal{E}} = \frac{Q_0^2}{2}\int_{R_+} ^{R^-} \frac{dr}{r^2} =
  \frac{Q_0^2}{2}\frac{\Delta R}{R^{+} \ R^{-}} \,  , \label{Mel_int2}
\end{equation}
where $\Delta R \equiv R^- -R^+$. This expression coincides with the
energy of a spherical capacitor with charge $Q_0$. To estimate the increase in
gravitational mass caused by the presence of an electric dipole layer,
we first estimate the corresponding dipole charge using Gauss' law. Using the
expression for the electric field outside of a spherically symmetric charge
distribution we obtain
\begin{equation}
 Q_0 = 4\pi\epsilon_0 r^2 \times E(r) \, . \label{Q0_exp}
\end{equation}
Since we are only dealing with a very narrow region near the surface, the
metric functions can safely be assumed to be constant there, and thus
the metric corrections can be incorporated in the constant $Q_0$.

\begin{figure*}
\centering
\vspace{0.8cm}
\includegraphics[width=13.0cm,clip, trim = 0 0 2.0 2.0]{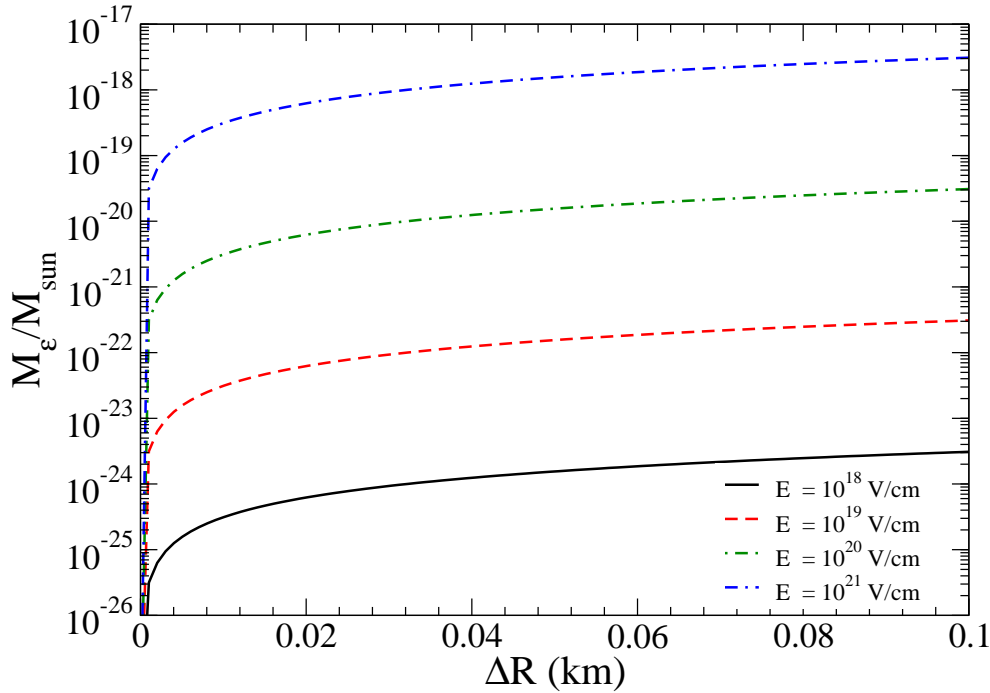}
\caption{\label{Dmel_graph} (color online) Increase of gravitational mass due to
  electric dipole energy, as a function of dipole width. The
  calculations are performed for a strange star with radius of 10 km
  and for the surface electric fields indicated.}
\end{figure*}

Using expression (\ref{Mel_int2}) and (\ref{Q0_exp}), and upon
performing the appropriate unit transformation we obtain
\begin{equation}
  \frac{M_{\cal{E}}}{M_{\odot}} =  3.12 \time 10^{-61}\  R_s^4 \ E^2 \
\frac{\Delta R}{R^+ \ 
    R^-}    \, , \label{Mel_final}
\end{equation}
with $E$ in V/cm and $R_s$, $\Delta R$, $R^+$ and $R^-$ in km. In Fig.\
\ref{Dmel_graph} we plot Eq.\ (\ref{Mel_final}) for a strange
star with 10 km radius and with the surface electric fields indicated.
Figure \ref{Dmel_graph} indicates that the increase in gravitational
mass of strange stars, resulting from the energy associated with the electric
dipole layer on the surface, is very small. We can conclude
then that
only when the star possess a net charge one might find a
significant contribution to the gravitational mass, due to a macroscopic
extension of the electric field outside of the star.

\section{Possible Mechanism of Magnetic Field Generation}
\label{section:IV}

In Section \ref{section:III}
we calculated the increase in gravitational mass that is caused by the
energy density of the surface dipole layer. These results are
calculated for a static strange star. In this section we will extend
this study to dipole fields on rotating strange stars.

\begin{figure}
\centering
\vspace{1.0cm}
\includegraphics[width=6.0cm]{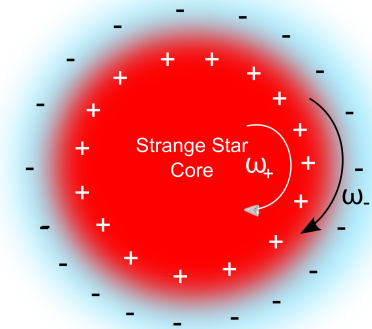}
\caption{\label{surf_current_scheme} (color online) Schematic
  illustration of the formation of electric currents at the surface of
  a rotating strange star. $\omega_+$ and $\omega_-$ are the
  frequencies at which the core and electron layer are rotating,
  respectively.}
\end{figure}

The most obvious effect that needs to be considered when taking
rotation into account is the formation of a magnetic field. Let us assume that
the star is rotating around the $z$ axis. Thus, if the
stellar core and the exterior electron layer are rotating at different
velocities, an electric current at the surface of the star will be created,
which leads to the formation of a magnetic field.  This situation is
schematically illustrated in Fig.\ \ref{surf_current_scheme}.
The electric surface current can be expressed as
\begin{equation}
 I = \sigma \, ( \omega_+ - \omega_-) \, ,
\end{equation}
where $\sigma$ is the surface charge density and $\omega_+$ and
$\omega_-$ are the frequencies at which the core and electron layer
are rotating, respectively. Obviously, if the core and electron layer are
rotating at the same frequency the surface currents are zero and no
magnetic field is formed.

The surface charge density can be estimated by using the result of
Section \ref{section:III}. The positive charge of the core of a strange
star has been estimated in Eq.\ (\ref{Q0_exp}). Using this result, we
can express $\sigma$ as
\begin{equation}
  \sigma = \frac{Q_0}{A} = \frac{4\pi\epsilon_0 R^2 E(R)}{4\pi R^2} = 
  \epsilon_0 \,  E(R) \, ,
\end{equation}
with $E(R)$ the electric field at the surface of strange stars (predicted for
the case where the core and electron layer are rotating at the
same frequency).

\begin{figure*}
\centering
\includegraphics[width=13.0cm]{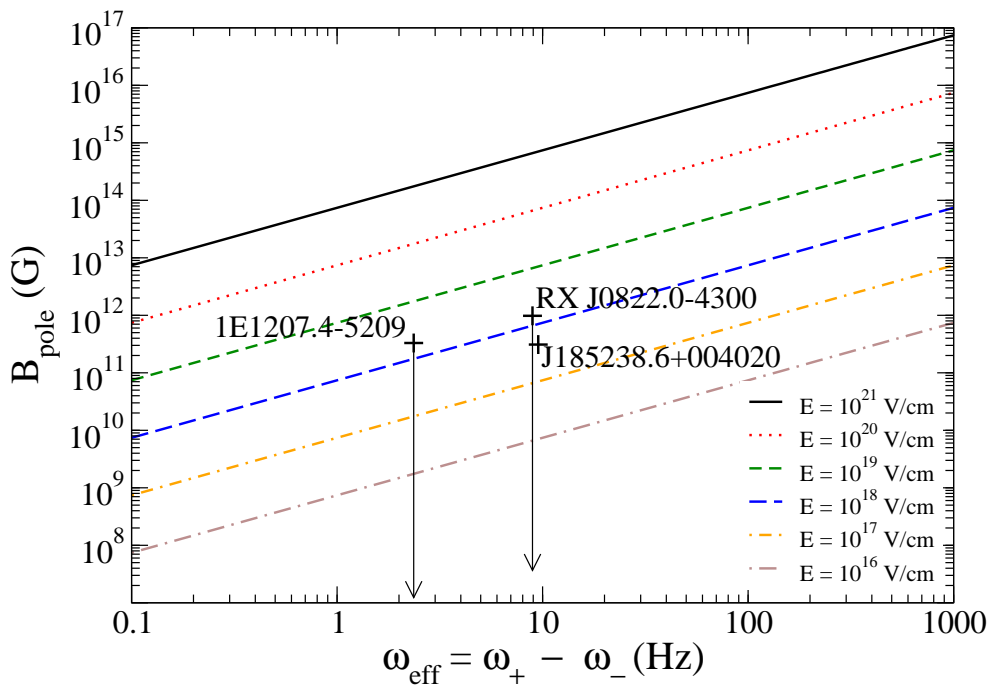}
\caption{\label{B_pole_graph} (color online) Polar magnetic field for strange
quark
  stars as a function of the difference in frequency between the
  positively charged core and negatively charged electron
  layer. Different electric fields indicate the value of the static
  electric field the star would have in the case that both the core
  and the electron layer are rotating at the same frequency
  ($\omega_{\text{eff}} = 0$). Also plotted is the observed magnetic
  field and period for 3 Central Compact Objects (CCO)
  (\cite{Halpern2010} and references therein)}
\end{figure*}

The magnetic field of a rotating, electrically charged sphere can be
easily calculated \cite{Jackson}. The result is a uniform magnetic
field pointing in the $+\hat{z}$ direction for the
  inside of the sphere, and a dipole field for the outside. The exterior field
can then be written as
\begin{equation}
  \vec{B} = \frac{1}{3} \, \mu_0 \, \sigma \, (\omega_+ - \omega_-) 
  \frac{R^4}{r^3}(2 \cos  \theta \, \hat{r} + \sin \theta \, \hat{\theta}) 
  \, .
\end{equation}
Where $\theta$ is the polar angle, $\hat{r}$ and $\hat{\theta}$ are unit
vector for the radial and polar directions.
After the appropriate unit conversions we can write the magnetic field at the
pole and at the equator as
\begin{eqnarray}
  B_{\text{p}} = E (\omega_+ - \omega_-) R \times 7.4104 \times 10^{-9} ~ 
  \text{G} \, , \\
  B_{\text{eq}} = E (\omega_+ - \omega_-) R \times 3.7052 \times 10^{-9} ~
  \text{G} \, , 
\end{eqnarray}
with $R$ in km, $E$ in V/cm and $\omega$ in Hz. It is important to note that
the predicted magnetic field is proportional to both the surface electric
field, and the effective rotation frequency. In Fig.\
\ref{B_pole_graph} we show the polar magnetic field as a function of
the effective frequency ($\omega_{\text{eff}} = \omega_+ - \omega_-
$), for strange quark stars with different electric fields.
Figure \ref{B_pole_graph} shows that for typical frequencies expected for
compact stars, the resulting magnetic
field can be rather high. For the extreme case with
effective frequency $\sim 700$ Hz, the magnetic field might be as
high as $10^{15}$ G, for static electric fields of $10^{19}$ V/cm.
Magnetic fields of this magnitude are expected for magnetars. These
objects, however, have relatively low rotational frequencies, and
therefore an effective frequency of the order of $700$ Hz is
unlikely. For a more moderate effective frequency of say 10 Hz, one
obtains magnetic fields in the range of $B \sim 10^{10} - 10^{11}$~G. Such
$B$ values and frequencies are in agreement with observations made for
some Central Compact Objects (CCO). These objects are characterized by
a steady flux predominately in the X-ray range and the lack of optical
and radio counterparts. There have been observations of the magnetic
fields and frequencies for three of these objects (see
\cite{Halpern2010} and references therein). Their quantities are
listed in Table \ref{table:CCO} and indicated in Fig.\
\ref{B_pole_graph}.
\begin{table}[htb]
  \caption{\label{table:CCO} Observed magnetic fields and frequencies of
    three Central Compact Objects (CCOs). (Data from \cite{Halpern2010}
    and references therein.)}
\begin{ruledtabular}
\begin{tabular}{ccc}
  CCO & $\Omega$ (Hz)& $B$ ($10^{11}$ G)\\
  \hline
  RX J0822.0-4300 &  8.928  & $< 9.8$  \\
  1E 1207.4-5209  &  2.3584 & $< 3.3$  \\
  CXOU J185238.6+004020  &   9.5238  & $3.1$
\end{tabular}
\end{ruledtabular}
\end{table}
Assuming that the observed frequency of these objects equals the
effective differential frequency (i.e.\ the difference between the
frequencies of the core and electron layer), we find that the
predicted magnetic fields are in good agreement with the observed
ones. Figure \ref{B_pole_graph} shows that stars whose static electric
fields are up to $10^{18}$ V/cm can generate the magnetic fields
observed for some CCOs. This finding may indicate that these objects
may be strange stars rather than neutron stars. More detailed studies
are necessary, however, to put this conjecture on a firm basis.

\section{A possible heating mechanism}
\label{section:heating}

As discussed in Section \ref{section:IV}, we allow the strange quark
core and the electron layer to be differentially rotating. Considering
that the electron layer is separated from the core by distances of the
order of 1000 fm, one should expect some frictional heating in this
region. A similar scenario is discussed in references
\cite{Shibazaki1989,VanRiper1995}, where a layer of superfluid matter
at the inner crust is differentially rotating with respect to the rest
of the star with average differential frequency between 0.1 and 5 Hz. Following
the footsteps of references
\cite{Shibazaki1989,VanRiper1995}, we estimate the frictional heating
for our model, in which the electron layer is differentially rotating
with respect to the strange star core. For this we hereafter assume
that the core is at rest ($\omega_+ = 0$) with respect to the electron
layer, which is rotating with a frequency of $\omega_- = 10$ Hz. The
frictional heating is given by the difference between the change in
rotational energy, and the rate at which the torque that is causing
the layer to spin down, is doing work,
\begin{equation}
 H = -\frac{d}{dt}\left( \frac{I_- \omega_-^2}{2}\right) - \arrowvert \tau
\arrowvert \omega_-. \label{heating}
\end{equation}
Where $H(t)$ is the heating, $I_-$ is the moment of inertia of the electron
layer, and $\tau$ is the torque, which is given by
\begin{equation}
\tau = I_-  \dot{\omega}_- \, . \label{torque}
\end{equation}
In Eq. (\ref{torque}), $\dot{\omega}_-$ is the spin-down frequency for the
electron layer.
Combining Equation (\ref{torque}) with (\ref{heating}) we get the following
expression for the heating
\begin{equation}
 H(t) = 2 I_-  ~ \omega_- ~ \arrowvert\dot{\omega}_-\arrowvert.
\end{equation}

We use the Newtonian approximation to calculate the moment of inertia of a
spherical shell, and using the observed properties for object 
J1852 \cite{Halpern2010}, which are $\dot{\omega} =-7.88\times 10^{-16}$
s$^{-2}$ and $\omega = 9.5238$ Hz, we get the following heating rate
\begin{equation}
 H = 5.279 \times 10^9 ~ \text{erg s}^{-1} \, .
\end{equation}
 
We are now in the position to estimate the time it would take for the
electron layer to dissipate through friction all of its rotational
energy. Assuming a constant spin-down rate, we find that the time
needed to convert the total rotation energy of the layer into heat is
$\sim 5.21 \times 10^{7}$ years, which is the same time scale of the cooling of
a compact star. This figure constitutes an
order-of-magnitude estimate only. Carrying out detailed microscopic
calculations (see \cite{Petrovich2010}, for example) on the re-heating
of strange stars by differentially rotating electron spheres (and the
associated time scales) is beyond the scope of this work, but will be
carried out in a future study.


\section{Role of color superconductivity}
\label{section:superconducting}

In the discussions of Sections \ref{section:II}, \ref{section:III} and
\ref{section:IV}, the possibility of pairing in quark matter was not addressed.
If strange quark matter stars exist, they are probably in a superconducting
state. The plausible condensation pattern of such matter at densities $\gg
\rho_0$ ($\rho_0$ being the nuclear saturation density) is the CFL
phase \cite{Alford2001}. The interior of stellar CFL matter, which is subject
to the conditions of chemical equilibrium and electric charge neutrality, is
characterized by equal numbers of u, d, and s quarks and, thus, the
total absence of electrons. The situation is different at the surface
of stellar CFL matter \cite{Madsen2001, Usov2004}
where the number of s quarks is reduced
with respect to u and d quarks because of changes in the density of
states. This leads to the presence of electrons at the surface of CFL
matter and the formation of an electric dipole layer at the surface of
a CFL strange quark star. (Usov has shown \cite{Usov2004} that the electric
field generated at the surface of a CFL
strange star is even higher than the one at the surface of a
non-superconducting strange quark matter, reaching values of $\sim 10^{18}$ 
V/cm.)

For intermediate densities ($\sim 2\rho_0$) the condensation pattern of
strange quark matter is less clear. Model calculations indicate that
for such densities strange quark matter may be in the 2SC phase
\cite{Alford2008} where only the u and d quarks of two colors are paired. In
this case
electrons will be present throughout the stellar CFL strange quark
matter to maintain electric charge neutrality. The same is the case if
strange quark matter were to form a crystalline color superconductor
\cite{Alford2001a,Bowers2002}. In the
latter event the momenta of the quark pairs do not add up to zero,
requiring the presence of electrons in such matter as well.

The bottom line of all this is that, independent of the specific
condensation pattern of color superconducting strange quark matter,
there will always exist electron dipole layers at the surfaces of CFL
strange quark matter stars, whose physical consequences are discussed
in this paper.

\section{Conclusions}\label{section:conclusions}

In this paper we have analyzed the surface properties of bare strange
stars, focusing on their electromagnetic properties. We extended the
work presented in \cite{Bekenstein1971,Negreiros2009,Ray2003}, where
the bulk properties of compact stars possessing a net electric charge
were investigated. Here we consider a bare strange star with zero net
electric charge. For such stars, as already shown in
\cite{Alcock1986,Alcock1988,Usov2004}, because of the displacements of
electrons, the stellar quark core becomes positively charged and the
region outside of it becomes negatively charged, leading to an
electric dipole layer. We used the general relativistic stellar
structure equations of electrically charged compact stars
\cite{Bekenstein1971,Negreiros2009,Ray2003} to calculate the increase
in gravitational mass that originates from the energy of the electric
dipole. We found that even for macroscopic (unrealistically large)
dipole widths (on the order of a 100 m), the increase in
gravitational mass is negligible ($\sim 10^{-20}~M_{\odot}$). We can
thus safely conclude that only strange stars which possess net electric
charges, as considered in \cite{Negreiros2009}, may lead to distinct
increases in gravitational mass. 

The second part of this paper deals with electromagnetic effects at
the surfaces of electrically charged strange stars, which emerge if
the star and the electron sphere surrounding the star should rotate at
different frequencies.  In this event electric currents would be
created at the stellar surfaces of the star. The strength of these
currents is determined by the magnitude of the net electric charge
available and by the difference in the rotational frequencies of the
stellar core and the electron layer.  The magnetic field of such a
configuration was found to be
uniform inside the star, and a dipole type outside. We also found that,
depending on the electric
field and the relative (effective) frequency between the star and the
electron layer, the magnetic field may be as high as
$10^{16}$~G. Such a strong magnetic field can only be achieved for very
high static electric fields on the order of $\sim 10^{20}-10^{21}$~V/cm and
effective frequencies of $\sim 700 - 1000$~Hz. For small effective
rotational frequencies of say $\sim 10$~Hz and more moderate static
electric fields of $\sim 10^{16}-10^{18}$~V/cm one obtains magnetic fields
on the order of $10^{9}-10^{11}$~G. This is a very intriguing result
because such magnetic field values and rotational frequencies are in
good agreement with the observed magnetic fields and frequencies of
three Central Compact Objects (CCO).  
 These objects have
relatively long rotational periods, and for the three cases for which
data exists, small magnetic fields of $\sim
10^{11}$~G \cite{Halpern2010}. These 
objects could thus be comfortably 
interpreted as strange stars whose electron atmosphere rotates at a 
frequency that is slightly
different from the strange star. It is important to stress that the model
proposed by us establishes a connection between CCO's and strange
quark matter stars, that may possibly be in the CFL superconducting phase. There
is also the possibility that the strange star presents chromomagnetic
instabilities, if the matter is in a 2SC phase \cite{Alford2008}. This issue
needs to be addressed carefuly, and we will
leave this topic for future investigations.

\begin{acknowledgments} 
We acknowledge the access to the computing facility of the Center of Scientific
Computing at the Goethe-University Frankfurt for our numerical calculations.
This work was supported by the National Science Foundation under
Grants PHY-55873A. This work was also supported in part by the DFG grant 436
RUS 113/957/0-1 and grants NS-7235.2010.2 and RFBR 09-02-91331 (Russia).
\end{acknowledgments}

\end{document}